\documentclass[twoside,twocolumn,english,prl]{revtex4}
\usepackage[]{fontenc}
\usepackage[latin1]{inputenc}
\usepackage{amsmath}
\usepackage{amssymb}

\makeatletter

\providecommand{\LyX}{L\kern-.1667em\lower.25em\hbox{Y}\kern-.125emX\@}

\usepackage{graphicx}
\usepackage{txfonts}
\usepackage{hyperref}

\usepackage{babel}
\makeatother
\begin{document}

\title{Radiation-Induced {}``Zero-Resistance State'' and the Photon Assisted
Transport}

\author{Junren Shi$^{1}$ and X.C. Xie$^{2,\, 3}$}

\affiliation{$^{1}$Condensed Matter Sciences Division, Oak Ridge National Laboratory,
Oak Ridge, Tennessee 37831\\
 $^{2}$Department of Physics, Oklahoma State University, Stillwater,
Oklahoma 74078\\
 $^{3}$International Center for Quantum Structures, Chinese Academy
of Sciences, Beijing 100080, China}

\begin{abstract}
We demonstrate that the radiation induced {}``zero-resistance state''
observed in a two-dimensional electron gas is a result of the non-trivial
structure of the density of states of the systems and the photon assisted
transport. A toy model of a structureless quantum tunneling junction
where the system has oscillatory density of states catches most of
the important features of the experiments. We present a generalized
Kubo-Greenwood conductivity formula for the photon assisted transport
in a general system, and show essentially the same nature of the transport
anomaly in a uniform system. 
\end{abstract}
\maketitle
The recent discovery of the {}``zero-resistance state'' in a two-dimensional
(2D) electron gas (2DEG) presents a surprise to the physics community~\cite{Mani2002,Zudov2002,Zudov2001,Durst2003,Andreev2003,Anderson2003}.
In these experiments~\cite{Mani2002,Zudov2002,Zudov2001}, the magneto-resistance
of a 2DEG under the influence of a microwave radiation exhibits strong
oscillations vs. magnetic field. Unlike the well known Shubnikov-de
Hass oscillation, the period of such oscillation is determined by
the frequency of the microwave radiation, and the resistance shows
minima near $\omega /\omega _{c}=n+1/4$, where $\omega $ is the
frequency of the microwave radiation and $\omega _{c}$ is the cyclotron
frequency of electron in the magnetic field. When the microwave radiation
is strong enough, the {}``zero-resistance states'' are observed
around the resistance minima. Durst et al. proposed a theory~\cite{Durst2003}
which successfully explains the period and the phase of the magneto-resistance
oscillation and also yields the negative resistance at the positions
where the {}``zero-resistance state'' was observed in the experiments.
Andreev et al.~\cite{Andreev2003} pointed out that such negative
resistance state is essential to understanding the {}``zero-resistance
state'' , because the negative resistance is unstable in nature and
could be interpreted as the {}``zero-resistance'' by the measurement
techniques employed in those experiments. A similar conclusion is
also reached in Ref.~\onlinecite{Anderson2003}. In essence, the
existence of the negative resistance state is crucial in the current
stage of theoretical understanding of the phenomenon. 

In this Letter, we show that such negative resistance state is the
result of the non-trivial structure of the density of states of the
2DEG system and the photon assisted transport. The similar effect
of photon assisted transport could be observed in other systems. A
generalized Kubo-Greenwood formula is presented to provide a formal
theory for such phenomena. 

To demonstrate our point in a clear and simple way, we first consider
the transport through a quantum tunneling junction. Such a toy model
catches most of the qualitative feature of the 2DEG experiments~\cite{Zudov2001,Mani2002,Zudov2002}.
At the same time, the simplicity of the model provides us a clear
view to the origin of the transport anomaly. Then we will present
a generalized Kubo-Greenwood formula to calculate the conductivity
of a general system under the influence of radiation, and provide
a natural explanation of the success of the simple toy model. 

\begin{figure}[tbph]
\includegraphics[  width=0.80\columnwidth]{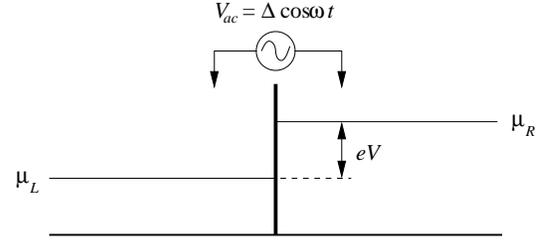}

\caption{\label{figure:structure}Toy model of a quantum tunneling junction.
A DC voltage $V$ and an AC field $V_{\mathrm{ac}}=\Delta \cos \omega t$
are applied on a structureless tunneling junction. The AC field models
the microwave radiation. }
\end{figure}

The structure of the toy model is shown in Fig.~\ref{figure:structure}.
We define the two systems across the junction as the left and right
leads. The AC voltage across the junction, $V_{\mathrm{ac}}=\Delta \cos \omega t$,
models the microwave radiation in a realistic experiment. The current
through the junction can be written as~\cite{Tien1963},\begin{eqnarray}
I & = & eD\int d\epsilon \sum _{n}J_{n}^{2}\left(\frac{\Delta }{\hbar \omega }\right)\left[f(\epsilon )-f(\epsilon +n\hbar \omega +eV)\right]\nonumber \\
 &  & \times \rho _{L}(\epsilon )\rho _{R}(\epsilon +n\hbar \omega +eV).\label{eq:jtot}
\end{eqnarray}
 where $\rho _{L(R)}$ is the density of states of the left (right)
lead, $f(\epsilon )$ is the Fermi distribution function, $D$ is
the constant in proportion to the transmission coefficient of the
junction, and $J_{n}(x)$ is the Bessel function of $n$th order.
The terms with $n\neq 0$ result from the photon-assisted tunneling
process. It has been assumed in Eq.~\ref{eq:jtot} that the electrons
tunneling across the junction are immediately removed from the junction
by the electric field in the leads so that there is no significant
charge accumulation in the junction area. This usually requires both
leads to be good conductors. However, as we will show later, in the
abnormal transport regime where the conductance becomes negative,
the assumption breaks down. 

For simplicity, we consider a symmetric system, $\rho _{L}(\epsilon )=\rho _{R}(\epsilon )=\rho (\epsilon )$.
In this case, the zero-bias conductance $\sigma =dI/dV|_{V=0}$ can
be written as, \begin{eqnarray}
\sigma  & = & e^{2}D\int d\epsilon \sum _{n}J_{n}^{2}\left(\frac{\Delta }{\hbar \omega }\right)\left\{ \left[-f^{\prime }(\epsilon )\right]\rho (\epsilon )\rho (\epsilon +n\hbar \omega )\right.\nonumber \\
 &  & \left.+[f(\epsilon )-f(\epsilon +n\hbar \omega )]\rho (\epsilon )\rho ^{\prime }(\epsilon +n\hbar \omega )\right\} .\label{eq:conductivity}
\end{eqnarray}
 The first term of the equation is the usual contribution from the
photon emission and absorption and is always positive. The second
term depends on the derivative of the density of states, and can be
either positive or negative. The contribution from the second term
is purely due to the photon-assisted tunneling process, and vanishes
when there is no AC field.

Based on Eq.~\ref{eq:conductivity}, it is not difficult to design
a system with the necessary form of the density of states to realize
a negative conductance. This is especially feasible for artificial
quantum systems~\cite{Keay1995}. However, to make our following
discussion more focused, we assume the density of states in the leads
is a periodic function of energy near the Fermi surface, and the period
in energy is $\hbar \omega _{c}$. By assuming that, we will show
that such a simple toy model catches most of the important features
of the experiments~\cite{Zudov2001,Mani2002,Zudov2002}.

Without invoking a special form for the density of states, we can
show that the conductance at the AC frequency $\omega =n\omega _{c}$
is identical to its dark field value, as observed in the experiments~\cite{Mani2002,Zudov2002,Zudov2001}.
In this case, $\rho (\epsilon +n\hbar \omega )=\rho (\epsilon )$,
the second term of Eq.~\ref{eq:conductivity} vanishes, leading to
\begin{eqnarray}
\sigma  & = & e^{2}D\int d\epsilon \sum _{n}J_{n}^{2}\left(\frac{\Delta }{\hbar \omega }\right)\left[-f^{\prime }(\epsilon )\right]\rho ^{2}(\epsilon )\nonumber \\
 & = & e^{2}D\int d\epsilon \left[-f^{\prime }(\epsilon )\right]\rho ^{2}(\epsilon )\equiv \sigma _{\mathrm{dark}}\, ,\label{eq:fixpoint}
\end{eqnarray}
 where we have used the identity $\sum _{n}J_{n}^{2}(x)=1$. 

\begin{figure}[t]
\includegraphics[  width=0.80\columnwidth]{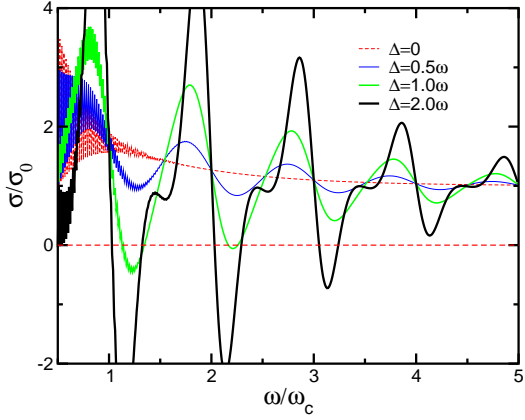}

\caption{\label{cap:conductance}Conductance dependence on $1/\omega _{c}$
for different radiation intensities. The parameters are the same as
those used in Ref.~\onlinecite{Durst2003}, $\mu =50\hbar \omega $,
$kT=0.25\hbar \omega $, $\omega \tau _{f}=6.25$. }
\end{figure}

We now assume the density of states has the following form, \begin{equation}
\rho (\epsilon )=\left(1+\lambda \cos \frac{2\pi \epsilon }{\hbar \omega _{c}}\right)\rho _{0}\, ,\label{eq:dos}\end{equation}
 where $\lambda $ is a dimensionless constant representing the fluctuation
amplitude of the density of states. A straightforward calculation
yields the conductance of the system,\begin{eqnarray}
\sigma (T)/\sigma _{0} & = & \sum _{n=-\infty }^{\infty }J_{n}^{2}\left(\frac{\Delta }{\hbar \omega }\right)\left[1+\frac{\lambda ^{2}}{2}\cos \left(2\pi n\frac{\omega }{\omega _{c}}\right)\right.\nonumber \\
 &  & \left.-n\pi \lambda ^{2}\frac{\omega }{\omega _{c}}\sin \left(2\pi n\frac{\omega }{\omega _{c}}\right)\right]+g\left(\frac{\mu }{\hbar \omega _{c}},T\right),\label{eq:sigmaT}
\end{eqnarray}
 where $\sigma _{0}=e^{2}D\rho _{0}^{2}$, and $g(\mu /\hbar \omega _{c},T)$
is the contribution from the Shubnikov-de Hass oscillation which diminishes
rapidly at finite temperatures. The conductance oscillation minima
can be easily determined from Eq.~\ref{eq:sigmaT}: for the $k$th
harmonics of the oscillation, the positions of the conductance minima
are given by the equation $\tan x=-x/2$, where $x=2\pi k\omega /\omega _{c}$.
For $k=1$, it yields the conductance minimum positions very close
to $\omega /\omega _{c}=n+1/4$, although not exactly. When the higher
orders of harmonics become important, we expect that the conductance
minima deviate from the $n+1/4$ rule. The amplitude of oscillation
is independent on the temperature, indicating any temperature dependence
observed in the experiments should come from the temperature dependence
of the density of states, \emph{i.e.,} $\lambda $. 

Next we use a more realistic density of states for the leads: when
the leads are the 2DEGs under a perpendicular weak magnetic field~\cite{Ando1982,Xie},
$\lambda $ is a function of $\omega _{c}$~\cite{Ando1982},\begin{equation}
\lambda =2\exp \left(-\frac{\pi }{\omega _{c}\tau _{f}}\right),\label{eq:lambda}\end{equation}
 where $\omega _{c}$ is the cyclotron frequency of electron and $\tau _{f}$
is the relaxation time of electron which depends on the scattering
mechanisms of the system and the temperature. The conductance for
such system is shown in Fig.~\ref{cap:conductance}. It is evident
that our model system, although very different and much simpler, shows
striking resemblances to the experimental observations~\cite{Mani2002,Zudov2001,Zudov2002}
and the more realistic calculation~\cite{Durst2003}. The conductance
minima are found at the positions near $\omega /\omega _{c}=n+1/4$
for the low and intermediate intensities of the AC field. When the
intensity becomes even higher, the multi-photon process sets in, presenting
the high order harmonic components to the conductance oscillation.
As in the experiments~\cite{Mani2002,Zudov2001,Zudov2002}, one can
see two sets of crossing points at $\omega /\omega _{c}=n$ and $\omega /\omega _{c}=n+1/2$,
where the conductances equal their dark field values. As we have shown
above, the former is a general property of the periodic density of
states, whereas the later will be destroyed by high intensities of
radiation, as shown in Fig.~\ref{cap:conductance}. 

\begin{figure}[tbp]
 \includegraphics[  width=0.80\columnwidth]{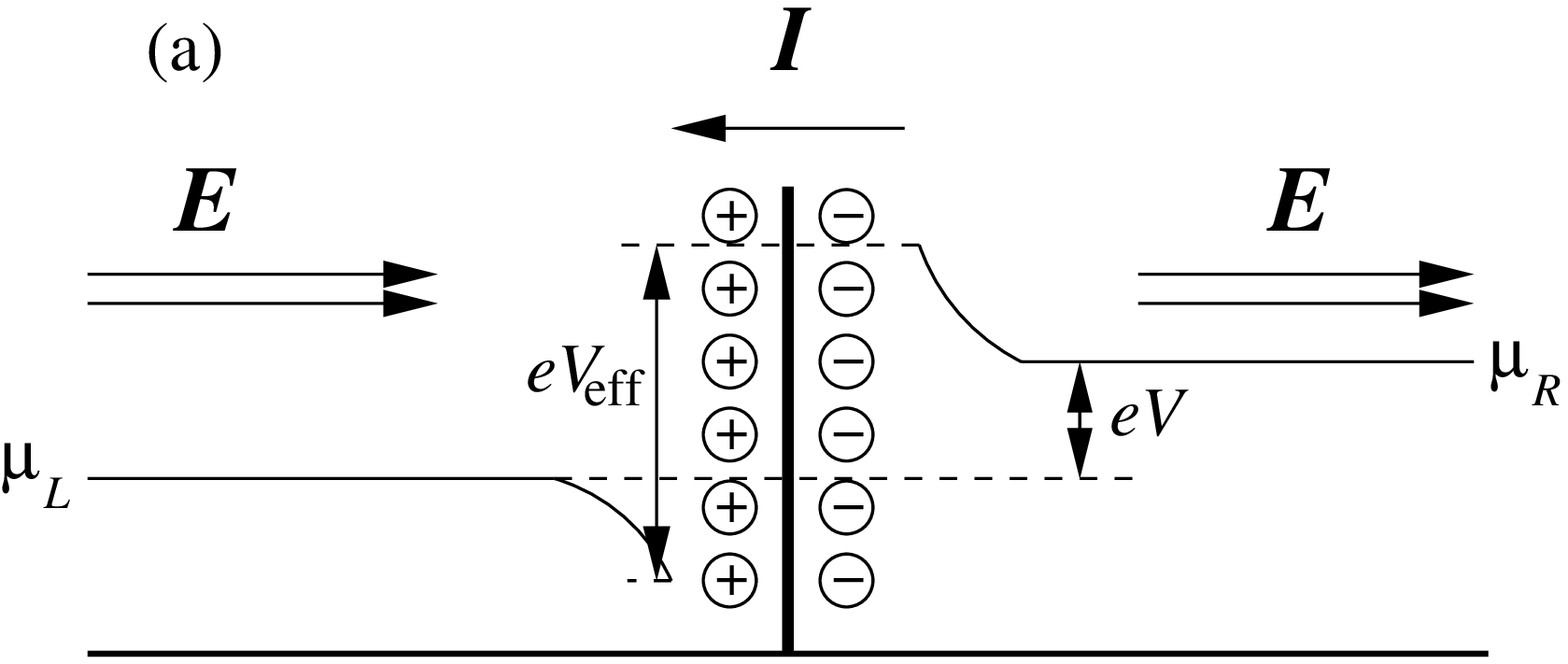}

\vskip 3mm 

\includegraphics[ width=0.40\columnwidth]{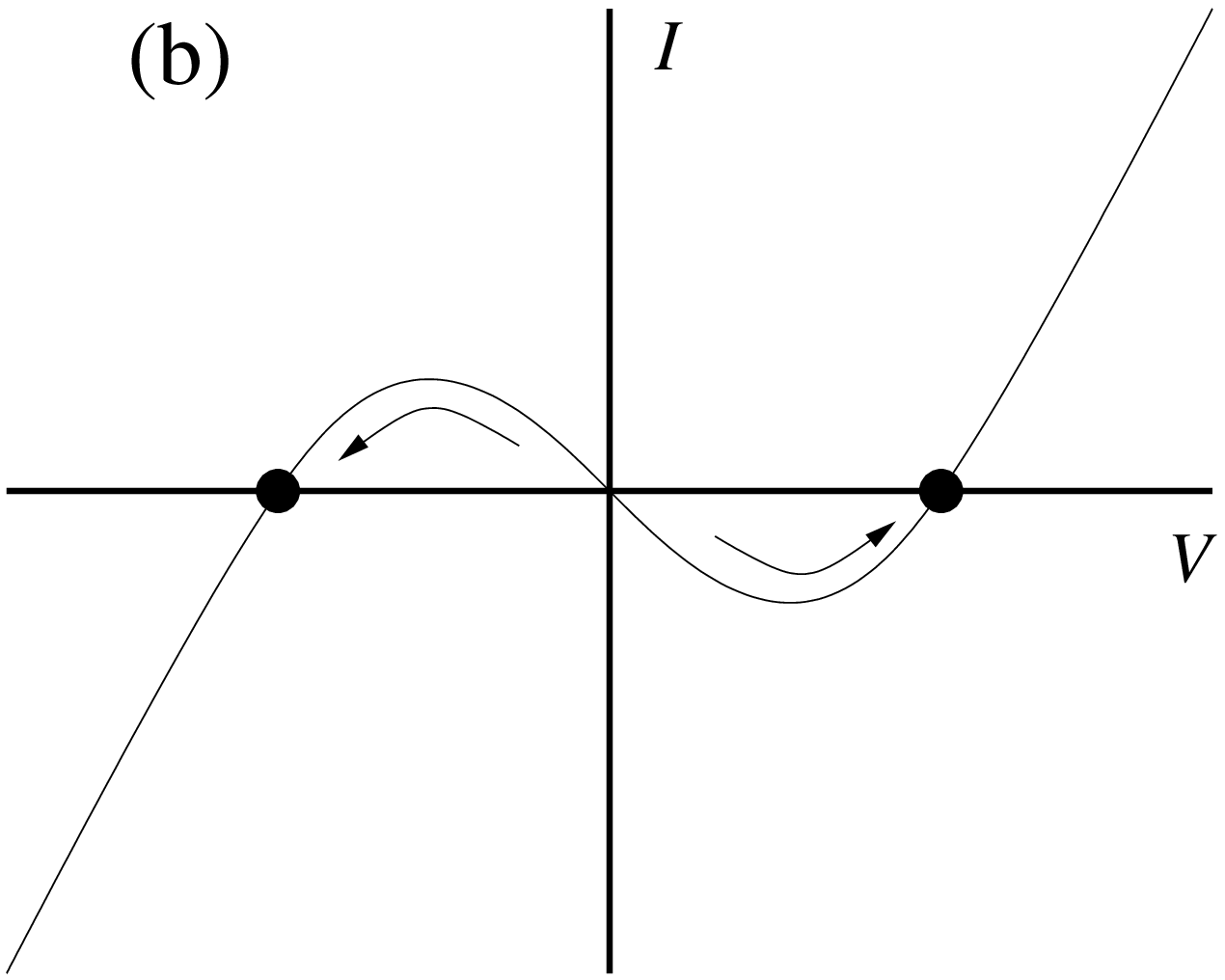}~\includegraphics[
width=0.40\columnwidth]{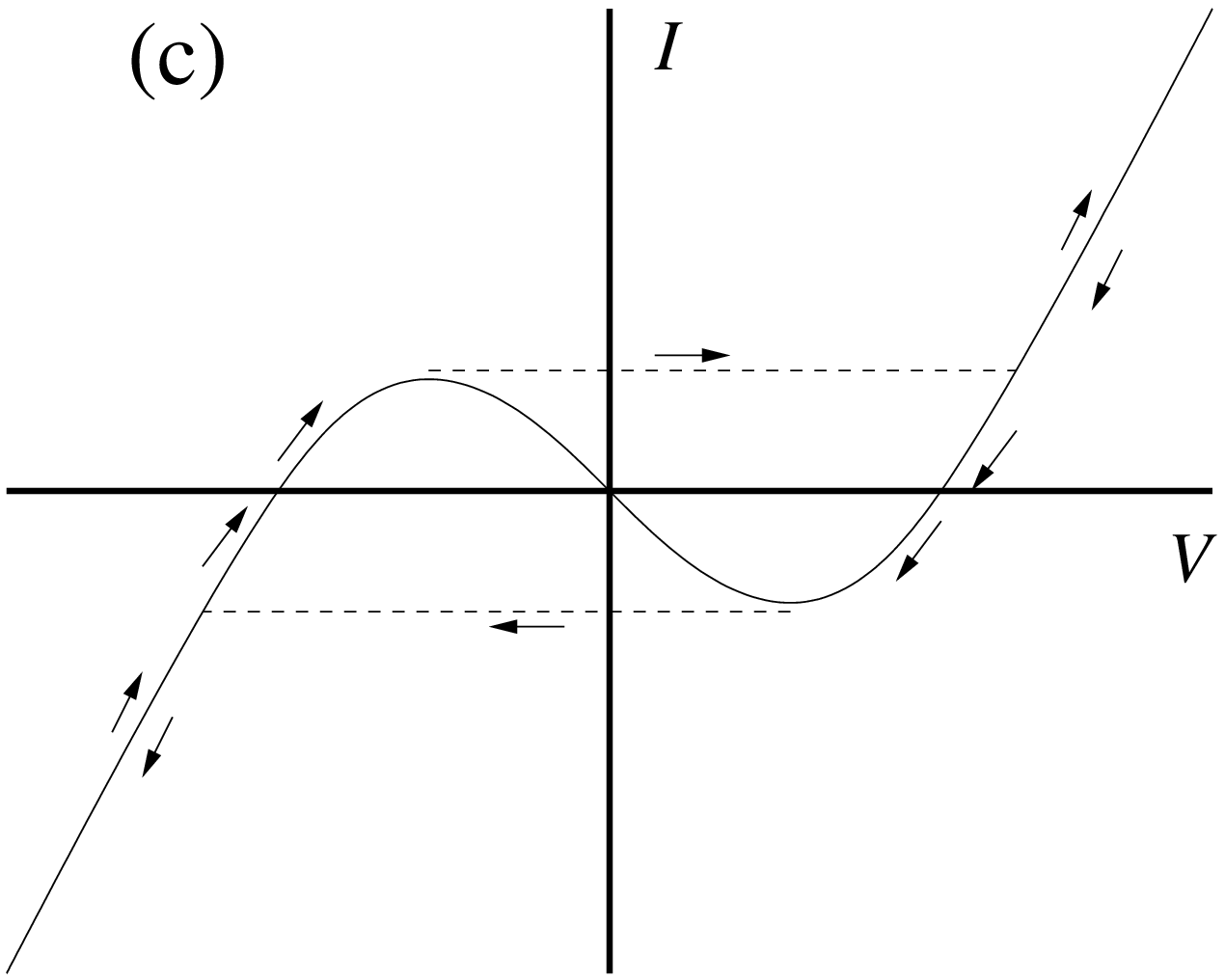}

\caption{\label{cap:negative-conductance}(a) Charge buildup in the negative
conductance regime at a constant voltage measurement. (b) Instability
in a constant voltage measurement. (c) Bi-stability in a constant
current measurement.}
\end{figure}

The system becomes unstable when entering into the negative conductance
regime. The consequence of such instability can be easily foreseen
in our toy model. In the case of constant voltage measurement, a negative
conductance means the current across the junction is in the reversed
direction to the electric field applied, as shown in Fig.~\ref{cap:negative-conductance}(a).
As a result, the tunneling electrons can not be removed from the junction
by the electric field in the leads. Instead, they accumulate near
the junction, and increase the effective voltage difference. The process
will continue on until the effective voltage difference between the
junction reaches such a point that the current becomes zero, as shown
in Fig~\ref{cap:negative-conductance}(b). Consequently, the measurement
will yield a zero conductance. On the other hand, in a constant current
measurement, the system will be in a bistable state with either positive
or negative junction voltage, depending on the history of the applied
current, as shown in Fig.~\ref{cap:negative-conductance}(c). We
stress that the analysis is very sensitive to the detailed setup of
the system. In the case of this toy model, many parameters such as
barrier thickness, lead configurations and dielectric constants may
affect the resulting phase. However, the instability itself is totally
determined by the radiation power and the density of states.

The photon-assisted transport process, which is responsible for the
transport anomaly in the tunneling junction, also exists in a uniform
system like the 2DEG. This becomes clear when we look at the Kubo-Greenwood
conductivity formula~\cite{Greenwood1958}, where the total conductivity
is a summation over all possible {}``tunneling'' between single
electron states. A generalization of the Kubo-Greenwood formula (see
Eq.~\ref{eq:sigma}) shows the similar contribution of the photon-assisted
tunneling. The effective AC voltage $\Delta _{\mathrm{eff}}$ in such
{}``tunneling'' is determined by the spatial separation between
the involved single electron states, and on average is of the order
of $E_{\omega }l$, where $l$ is the mean free path of electron,
$E_{\omega }$ is the radiation field strength. Based on the parameters
given in the experiment~\cite{Mani2002}, we deduce that $l\sim 10^{-4}\, \mathrm{m}$,
$E_{\omega }\sim 10\, \mathrm{V}/\mathrm{m}$, thus $\Delta _{\mathrm{eff}}\sim 1\, \mathrm{meV}$,
which is the same order of the radiation frequency $\hbar \omega \sim 0.4\, \mathrm{meV}$.
The estimation indicates the photon-assisted process is sufficient
to understanding the observed transport anomaly. The special experimental
setup, \emph{i.e.}, a 2D Hall system, is not essential, although the
observation does require the oscillatory density of states and a high
mobility sample to provide a sufficient long electron mean free path. 

Now we derive a generalized Kubo-Greenwood formula for the system
under the influence of radiation. For a general system considered,
the Hamiltonian can be written as, \begin{equation}
H=H_{0}+H_{\mathrm{ac}}(\omega )+H_{\mathrm{dc}}\, .\label{eq:Hamiltonian}\end{equation}
 Here $H_{0}$ is the Hamiltonian of the unperturbed system including
the disorder potential. $H_{\mathrm{ac}}(\omega )$ is the coupling
to the external radiation of frequency $\omega $. $H_{\mathrm{dc}}$
is the potential induced by a small DC field, and is treated as a
small perturbation to the system defined by $H_{1}=H_{0}+H_{\mathrm{ac}}$.
The standard linear response formula yield the current density through
the system as~\cite{Mahan},\begin{eqnarray}
\mathbf{J}(t) & = & \lim _{\omega _{0}\rightarrow 0}\frac{2\mathbf{E}_{0}}{\hbar \omega _{0}}\int _{-\infty }^{t}dt^{\prime }\left\langle \left[\hat{\mathbf{j}}(t),\hat{\mathbf{j}}(t^{\prime })\right]\right\rangle e^{-i\omega _{0}t^{\prime }+\eta t^{\prime }},\label{eq:linearresponse}
\end{eqnarray}
 where the DC field is simulated by an electric field $\mathbf{E}_{0}e^{-i\omega _{0}t+\eta t}$
with the infinitesimal frequency $\omega _{0}$, and $\hat{\mathbf{j}}=e/2\sum _{i}[\mathbf{v}_{i}\delta (\mathbf{x}-\mathbf{x}_{i})+\delta (\mathbf{x}-\mathbf{x}_{i})\mathbf{v}_{i}]$
is the current operator. We have omitted the gauge term which has
no contribution to the DC current. For brevity, we will drop $\lim _{\omega _{0}\rightarrow 0}$
in the following derivations. An extra factor of $2$ has been added
to address the spin degeneracy. 

Unlike the usual DC transport system, the unperturbed system $H_{1}$
is still time dependent. According to the Floquet theorem, the time
dependent wave function of such a system can be written as, \begin{eqnarray}
\left|\alpha (t)\right\rangle  & = & e^{-i\tilde{E}_{\alpha }t/\hbar }\sum _{n=-\infty }^{\infty }e^{-in\omega t}\left|\alpha ,n\right\rangle ,\label{eq:floquet}
\end{eqnarray}
 where $\tilde{E}_{\alpha }$ is quasi-energy of the Floquet state.
We assume the external radiation is applied onto the system adiabatically,
and the system keeps the adiabaticity during the process~\cite{Hone1997},
so each Floquet state can be uniquely mapped to an eigenstate $\left|\alpha \right\rangle $
for the system without the radiation. In general, $E_{\alpha }\neq \tilde{E}_{\alpha }$,
where $E_{\alpha }$ is the energy of the state $\left|\alpha \right\rangle $.
$\left|\alpha ,n\right\rangle $ is the appropriate wave function
determined by the solution of the time-dependent system. 

Turning to the Heisenberg representation, and using the eigenstates
of $H_{0}$ as the basis, we can expand the current operator as, \begin{eqnarray}
\hat{\mathbf{j}}(t) & = & e^{i\tilde{H}_{0}t/\hbar }\left[\sum _{n=-\infty }^{\infty }\hat{\mathbf{j}}_{n}e^{-in\omega t}\right]e^{-i\tilde{H}_{0}t/\hbar }\nonumber \\
\hat{\mathbf{j}}_{n} & = & \sum _{m}\left|\alpha \right\rangle \left\langle \alpha ,m\right|\hat{\mathbf{j}}\left|\beta ,m+n\right\rangle \left\langle \beta \right|,\label{eq:currentexpansion}
\end{eqnarray}
 where $\tilde{H}_{0}$ is the quasi-energy operator defined by $\tilde{H}_{0}\left|\alpha \right\rangle =\tilde{E}_{\alpha }\left|\alpha \right\rangle $. 

Substituting Eq.~\ref{eq:currentexpansion} into Eq.~\ref{eq:linearresponse},
and following the usual process of the derivation of the Kubo formula~\cite{Mahan},
we have, \begin{eqnarray}
\sigma _{\mathrm{dc}} & = & \frac{2\pi }{V}\frac{\partial }{\partial \omega _{0}}\sum _{f,i}\sum _{n}\left(P_{i}-P_{f}\right)\left|\left\langle f\right|\hat{\mathbf{j}}_{n}\left|i\right\rangle \right|^{2}\nonumber \\
 &  & \times \delta \left(\hbar \omega _{0}+n\hbar \omega -\tilde{E}_{f}+\tilde{E}_{i}\right),\label{eq:sigmadc}
\end{eqnarray}
 where $P_{i(f)}=e^{-\beta E_{i(f)}}/Z$ is the probability of the
system at the state $\left|i\right\rangle $ ($\left|f\right\rangle $).
Note the probability is determined by $E_{i(f)}$, the state energy
of the system without the radiation, instead of the Floquet quasi-energy
$\tilde{E}_{i(f)}$. The discrepancy between the Floquet quasi-energy
$\tilde{E}$ and the energy $E$ of the original system can be ignored
as long as the radiation is not extremely strong. 

For the non-interacting system, we can simplify Eq.~\ref{eq:sigmadc}
and obtain the generalized Kubo-Greenwood formula,\begin{eqnarray}
\sigma _{\mathrm{dc}} & = & 2\pi \frac{\partial }{\partial \omega _{0}}\sum _{n}\int d\epsilon _{\alpha }\left[f(\epsilon _{\alpha })-f(\epsilon _{\alpha }+\hbar \omega _{0}+n\hbar \omega )\right]\nonumber \\
 &  & \times \overline{\left|\left\langle \beta \right|\hat{\mathbf{j}}_{n}\left|\alpha \right\rangle \right|^{2}}\rho (\epsilon _{\alpha })\rho (\epsilon _{\alpha }+\hbar \omega _{0}+n\hbar \omega ),\label{eq:PI}
\end{eqnarray}
 where $\overline{\left|\left\langle \beta \right|\hat{\mathbf{j}}_{n}\left|\alpha \right\rangle \right|^{2}}$
is the average over all possible initial and final states for the
transitions $\epsilon _{\alpha }\rightarrow \epsilon _{\alpha }+n\hbar \omega +\hbar \omega _{0}$.
We have assumed $\tilde{\epsilon }_{\alpha }\approx \epsilon _{\alpha }$
in the derivation. In this equation, the same contribution of the
photon-assisted process is evident by comparing it with Eq.~\ref{eq:conductivity}. 

To gain more insight to Eq.~\ref{eq:PI}, we study a limiting case
where the wavelength of the radiation is much longer than the spatial
extend of the electron wave function. As the result, the AC electric
field felt by the individual electron state can be approximately considered
as spatially independent. So for state $\left|\alpha \right\rangle $,
\begin{equation}
H_{\mathrm{ac}}^{\alpha }\approx \Delta \cos (\omega t-\mathbf{k}\cdot \mathbf{r}_{\alpha }),\label{eq:vac}\end{equation}
 where $\Delta =eE_{\omega }c/\omega $, and $\mathbf{r}_{\alpha }$
is the average center of the wave function. Now the Floquet state
can be obtained analytically, \begin{eqnarray}
\left|\alpha (t)\right\rangle  & \approx  & e^{-i\epsilon _{\alpha }t/\hbar }\sum _{n=-\infty }^{\infty }J_{n}\left(\frac{\Delta }{\hbar \omega }\right)e^{-in(\omega t-\mathbf{k}\cdot \mathbf{r}_{\alpha })}\left|\alpha \right\rangle .\label{eq:alphat}
\end{eqnarray}
 Comparing Eq.~\ref{eq:alphat} with Eq.~\ref{eq:floquet}, we conclude
$\left|\alpha ,n\right\rangle =J_{n}\left(\Delta /\hbar \omega \right)\exp (in\mathbf{k}\cdot \mathbf{r}_{\alpha })\left|\alpha \right\rangle $
and $\epsilon _{\alpha }=\tilde{\epsilon }_{\alpha }$. With Eq.~\ref{eq:currentexpansion},
it is straightforward to get, \begin{eqnarray}
\left\langle \alpha \right|\hat{\mathbf{j}}_{n}\left|\beta \right\rangle  & = & e^{in\mathbf{k}\cdot (\mathbf{r}_{\alpha }+\mathbf{r}_{\beta })/2}J_{n}\left(\frac{\Delta _{\alpha \beta }}{\hbar \omega }\right)\hat{\mathbf{j}}_{\alpha \beta },\label{eq:jn}
\end{eqnarray}
 where $\Delta _{\alpha \beta }=\Delta |\exp (i\mathbf{k}\cdot \mathbf{r}_{\alpha })-\exp (i\mathbf{k}\cdot \mathbf{r}_{\beta })|\approx e|(\mathbf{r}_{\alpha }-\mathbf{r}_{\beta })\cdot \mathbf{E}_{\omega }|$,
is the effective AC potential between two states, and $\hat{\mathbf{j}}_{\alpha \beta }=\left\langle \alpha \right|\hat{\mathbf{j}}\left|\beta \right\rangle $. 

As a result, the generalized Kubo-Greenwood formula can be written
as, \begin{eqnarray}
\sigma _{\mathrm{dc}} & = & \frac{\partial }{\partial \epsilon _{0}}\sum _{n}\int d\epsilon D_{n}(\epsilon ,\epsilon +n\hbar \omega )\nonumber \\
 & \times  & \left[f(\epsilon )-f(\epsilon +\epsilon _{0}+n\hbar \omega )\right]\rho (\epsilon )\rho (\epsilon +\epsilon _{0}+n\hbar \omega )\label{eq:sigma}
\end{eqnarray}
 where $D_{n}(\epsilon ,\epsilon +n\hbar \omega )=2\pi \hbar \overline{J_{n}^{2}(\Delta _{\alpha \beta }/\hbar \omega )\left|\hat{\mathbf{j}}_{\alpha \beta }\right|^{2}}$.
It is clear that the total conductivity can be considered as the summation
of photon assisted {}``tunneling'' conductance between the electron
states. This proves our previous qualitative argument. 

Finally, we discuss the implication of negative conductance. In general,
negative conductance signifies the instability of the driven system
by an external microwave radiation. Such instability may drive the
system to a far-from-equilibrium regime where nonlinear and self-organizing
effects dominate, resulting intriguing and rich phenomena~\cite{Prigogine1977}.
One possible phase of such kind has been proposed to understand the
{}``zero-resistance state''~\cite{Andreev2003,Anderson2003}. On
the other hand, we stress that these are two separate issues: (a)
origin of the instability; and (b) new phase induced by the instability.
While the instability is totally determined by the radiation power
and the density of states, the determination of new phase requires
more detailed knowledge of the system, as demonstrated in our toy
model. Our analysis also suggests the structure of new phase will
be very sensitive to specific setup of the system, and deserves more
careful studies.

In conclusion, we demonstrate the existence of the negative conductance
in a quantum tunneling junction under the influence of a radiation
field. We trace the origin of such transport anomaly to the non-trivial
structure of the density of states of the system and the photon-assisted
transport. A generalized Kubo-Greenwood conductivity formula is presented
to show essentially the same nature of the anomalies observed in tunneling
junctions and in 2DEG systems. We expect the similar photon assisted
transport phenomena could be observed in other systems. We also suggest
that the instability signified by the negative conductance could result
rich and interesting phenomena.

We thank Fuchun Zhang for bringing the problem to our attention and
Biao Wu for a discussion of quantum adiabaticity and critical reading
of the manuscript. We also acknowledge fruitful discussions with R.R.
Du, R.G. Mani, N. Read and A.F. Volkov. J.S. thanks Dr. Zhenyu Zhang
for constant support and mentorship. This work is supported by the
LDRD of ORNL, managed by UT-Battelle, LLC for the U.S. DOE (DE-AC05-00OR22725),
and by U.S. DOE (DE/FG03-01ER45687).

\end{document}